\documentclass{webofc}
\usepackage[varg]{txfonts}   
\newcommand{\WT}{{\rm WT}}
\newcommand{\HQSS}{{\rm HQSS}}
\newcommand{\SU}{\mbox{SU}}
\wocname{EPJ Web of Conferences}
\woctitle{Resonance Workshop at Catania}
\begin{document}
\title{Charmed baryonic resonances in medium}
%
%

\author{Laura Tolos\inst{1,2}\fnsep\thanks{\email{tolos@ice.csic.es}} 
}

\institute{Instituto de Ciencias del Espacio (IEEC/CSIC), Campus Universitat 
Aut\`onoma de Barcelona, Facultat de Ci\`encies, Torre C5, E-08193 Bellaterra 
(Barcelona), Spain
\and
 Frankfurt Institute for Advanced Studies, Johann Wolfgang Goethe University, Ruth-Moufang-Str. 1,
60438 Frankfurt am Main, Germany  }

\abstract{%
We discuss the behavior of dynamically-generated charmed baryonic resonances in matter within a unitarized coupled-channel model consistent with heavy-quark spin symmetry. We analyze the implications for the formation of $D$-meson bound states in nuclei and the propagation of $D$ mesons in heavy-ion collisions from RHIC to FAIR energies.
}
\maketitle
\section{Introduction}
\label{intro}

Quantum Chromodynamics (QCD) is the basic theory of the strong interaction. In the low-energy regime, QCD becomes a strongly-coupled theory, many aspects of which are not yet understood. An important effort has been invested in exploring the QCD phase diagram for high density and/or temperature. In fact, the study of matter under extreme conditions has become one of the main research activities of several experimental programs, from the ongoing LHC/CERN  (Switzerland) \cite{lhc} to the forthcoming FAIR  (Germany) \cite{fair} projects. Until now the studies have been concentrated in matter within the light-quark sector due to energy constraints. With the on-going and upcoming research facilities, the aim is to move from the light-quark domain to the heavy-quark one and to face new challenges where charm and new symmetries, such as heavy-quark symmetry, will play a significant role.

The primary theoretical effort is to understand the interaction between hadrons that incorporate the charm degree of freedom. With data coming from  CLEO, Belle, BaBar \cite{facility00} and other experiments, charmed hadronic states have received a lot of attention. Moreover, it is expected that in the upcoming years the FAIR project will provide new insights on charmed hadron spectroscopy \cite{fair}. The ultimate goal is to understand whether these states can be understood within the quark model picture and/or qualify better as dynamically generated states via hadron-hadron scattering processes.

There has been a tremendous success in describing experimental data on hadron spectroscopy by means of approaches based on coupled-channel dynamics. Particularly, unitarized coupled-channel methods have been used in the meson-baryon sector with charm content \cite{Tolos:2004yg, Tolos:2005ft, Lutz:2003jw, Lutz:2005ip, Hofmann:2005sw, Hofmann:2006qx, Lutz:2005vx, Mizutani:2006vq, Tolos:2007vh,  JimenezTejero:2009vq, Haidenbauer:2007jq, Haidenbauer:2008ff,  Haidenbauer:2010ch, Wu:2010jy, Wu:2010vk, Wu:2012md, Oset:2012ap,Liang:2014kra}, mostly motivated by the parallelism between the $\Lambda(1405)$ and the $\Lambda_c(2595)$. 

However, some of these models are not fully consistent with heavy-quark spin symmetry (HQSS) ~\cite{Isgur:1989vq}. HQSS is a proper QCD symmetry that appears when the quark masses become larger than the typical confinement scale. Thus, we have  developed a model that incorporates HQSS constraints \cite{GarciaRecio:2008dp,Gamermann:2010zz,Romanets:2012hm,GarciaRecio:2012db,Garcia-Recio:2013gaa,Tolos:2013gta}.

Nuclear medium modifications have been then incorporated in order to study how charmed baryonic resonances are modified in matter and, consequently,  to analyze the properties of charmed mesons in nuclear matter, and the influence of these modifications in the charmonium production rhythm at finite baryon densities. Possible variation of this rhythm might indicate the formation of the quark-gluon plasma (QGP) phase of QCD at extreme conditions. 

In this work we aim at investigating nuclear medium effects on dynamically-generated charmed baryonic resonances and the consequences for  the properties of open-charm mesons in dense nuclear matter and nuclei. We then test open-charm meson properties in matter by addressing the possible formation of $D$-mesic nuclei and  the $D$-meson propagation from FAIR to RHIC energies.

\section{Charmed baryonic resonances incorporating Heavy-Quark Spin Symmetry}
\label{spec}

One of the primary goals in the physics of hadrons is to understand the nature of newly discovered states, whether they can be described within the quark model picture and/or as hadron-hadron molecules. By adopting the latter approach,  we analyze recent developments in the description of experimental charmed baryonic states. 

Charmed baryonic states are dynamically generated by the scattering of mesons and baryons within a unitarized coupled-channel approach. We employ a model that explicitly incorporates HQSS~\cite{Isgur:1989vq}. HQSS predicts that, in QCD, all types of spin interactions involving heavy quarks vanish for infinitely massive quarks, thus, connecting vector and pseudoscalar mesons containing charmed quarks. Moreover,  chiral symmetry fixes  the lowest order interaction between Goldstone bosons and other hadrons by means of the Weinberg-Tomozawa (WT) interaction. Then, we aim at developing a predictive model for four flavors including all basic hadrons (pseudoscalar and vector mesons, and $1/2^+$ and $3/2^+$ baryons) which reduces to the WT interaction in the sector where Goldstone bosons are involved and which incorporates HQSS in the sector where charm quarks participate. Indeed, this is a model assumption that is justified in view of the reasonable outcome of the SU(6) extension in the three-flavor sector \cite{Gamermann:2011mq} and on a formal plausibleness on how the interaction in the charmed pseudoscalar meson-baryon sector comes out in the vector-meson exchange picture.

The extended WT model with HQSS constraints is given by
\begin{equation}
V = \frac{K(s)}{4f^2} H^\prime_\WT
,\qquad
H^\prime_\WT = H_{\rm ex}+H^\prime_{\rm ac} ,
\label{eq:2.21}
\end{equation}
with $K(s)$ a function of energy and $f$ the weak decay constant of mesons. The two mechanisms $H_{\rm ex}$ and $H^\prime_{\rm ac}$ are given in terms of quark and antiquark
propagation. The $H_{\rm ex}$ is the exchange term, in which a quark is
transferred from the meson to the baryon, as another one is transferred from
the baryon to the meson. The annihilation-creation $H^\prime_{\rm ac}$  mechanism takes into account that 
a light antiquark in the meson annihilates with a similar light quark
in the baryon, with subsequent creation of a light quark and a light antiquark.  The creation and annihilation of charmed
quark-antiquark pairs is removed in order to satisfy HQSS symmetry \cite{Garcia-Recio:2013gaa}.
This scheme fulfills some desirable requirements: i) it has symmetry
$\SU(6)\times \HQSS$, i.e., spin-flavor symmetry in the light sector and HQSS in the
heavy sector, the two invariances being compatible; ii) it reduces to SU(6)-WT
in the light sector, so it is consistent with chiral symmetry in that sector. 

With this potential we can solve the on-shell Bethe-Salpeter equation in coupled channels so as to calculate the scattering amplitudes \cite{Tolos:2013gta}. The poles of the scattering amplitudes are the dynamically-generated charmed baryonic resonances. 

Dynamically generated states in different charm and strange sectors  are
predicted within our model \cite{GarciaRecio:2008dp,Gamermann:2010zz,Romanets:2012hm}. We restrict our study to those states stemming from the most attractive representations of our SU(6) $\times$ HQSS model. Some of them can be
identified with known states from the PDG \cite{pdg}, by comparing the PDG data on these states with the mass,
width and, most important, the coupling to the meson-baryon channels of our
dynamically-generated poles. In this work, as an example, we present results in the $C=1,S=0, I=0$ and $C=1, S=0, I=1$ sectors. 

In the $C=1,S=0, I=0$ sector, three $\Lambda_c$ and one $\Lambda_c^*$ states are obtained \cite{Romanets:2012hm}. The experimental $\Lambda_c(2595)$
resonance can be identified with a pole that we obtain around 2618.8 MeV, as
similarly done in Ref.~\cite{GarciaRecio:2008dp}. A second broad $\Lambda_c$
resonance at 2617~MeV is, moreover, observed with a large coupling to the open
channel $\Sigma_c \pi$, very close to  the $\Lambda_c(2595)$, presenting the
same two-pole pattern found in the charmless $I = 0, S = -1$ sector for the
$\Lambda(1405)$ \cite{Jido:2003cb}. A third spin-$1/2$ $\Lambda_c$ resonance is
seen around 2828~MeV  and cannot be assigned to any experimentally known
resonance.  As for the case of spin-$3/2$ resonances, we find one at $(2666.6 -i 26.7 \ \rm{MeV})$ that is assigned to the experimental $\Lambda_c(2625)$. 

For $C=1, S=0, I=1$ ($\Sigma_c$ sector), three $\Sigma_c$ resonances are obtained with masses 2571.5, 2622.7 and 2643.4~MeV and widths 0.8, 188.0 and 87.0~MeV, respectively \cite{Romanets:2012hm}. Moreover, two spin-$3/2$ $\Sigma_c$ resonances are generated dynamically \cite{Romanets:2012hm}. The first spin-$3/2$, a bound state at 2568.4~MeV, is thought to be the charmed counterpart of the $\Sigma(1670)$, while the second state at $2692.9 -i 33.5$ ~MeV has not a direct experimental comparison. We can not assign any of these low-lying states to the experimental $\Sigma_c(2800)$ \cite{pdg}.

\section{Charmed baryonic resonances in matter and the open charm spectral functions}
\label{medium}

\begin{figure}
\begin{center}
\includegraphics[width=0.6\textwidth,height=7cm]{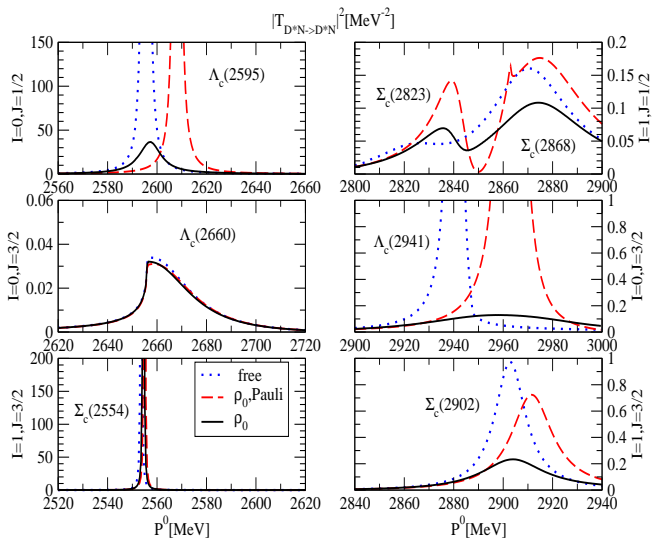}
\caption{Charmed baryonic resonances in dense matter (taken from Ref.~\cite{Tolos:2009nn}).}
\label{fig:reso}
\end{center}
\end{figure}

\begin{figure}
\begin{center}
\includegraphics[width=0.4\textwidth,height=5.5cm]{art_spec.eps}
\includegraphics[width=10cm, angle=90,height=5.5cm]{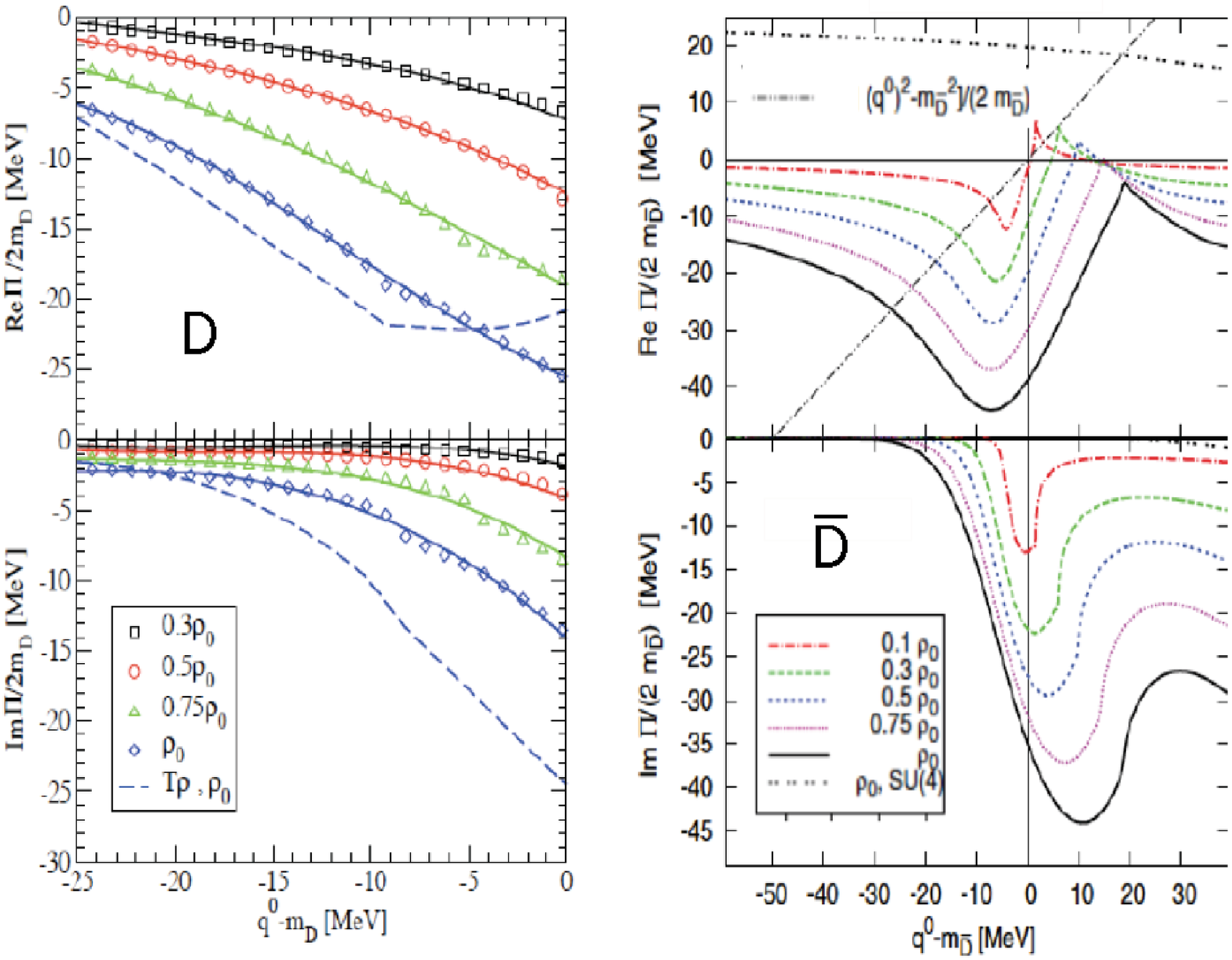}
\caption{Left:  The $D$ and $D^*$ spectral functions in dense nuclear matter at $\vec{q}=0$ MeV/c (taken from Ref.~\cite{Tolos:2009nn}). Right:  The $D$ and $\bar D$ optical potential at $\vec{q}=0$ MeV/c for different densities (taken from Refs.~\cite{GarciaRecio:2010vt,GarciaRecio:2011xt}). }
\label{fig2}
\end{center}
\end{figure}

The mass and width of the dynamically-generated states can be modified once these states are generated in a dense medium. In Fig.~\ref{fig:reso} we show the squared amplitude of $D^*N$-$D^*N$ transition for different partial waves as function of the center-of-mass energy for zero total momentum.  We analyze an energy range up to 3.5 GeV, much wider than the one studied in the previous section. Moreover,  we have fitted the position of the $\Lambda_c(2595)$ and, as a consequence, the masses of the other states are slightly modified as compared to the previous results in free space reported in Sec.~\ref{spec} . 

We find seven resonances predicted by the SU(6)xHQSS model that have or can have experimental confirmation:  i.e., ($I=0$, $J=1/2$) $\Lambda_c(2595)$,
($I=1$,$J=1/2$) $\Sigma_c(2823)$ and $\Sigma_c(2868)$, ($I=0$,
$J=3/2$) $\Lambda_c(2660)$,($I=0$, $J=3/2$) $\Lambda_c(2941)$, ($I=1$,
$J=3/2$) $\Sigma_c(2554)$ and ($I=1$, $J=3/2$) $\Sigma_c(2902)$
resonances. We analyze three different cases:  free space calculation, computation including Pauli blocking on the nucleonic
intermediate states at normal nuclear matter density $\rho_0=0.17
~ \rm{fm}^{-3}$, and  in-medium solution which
incorporates Pauli blocking effects and the $D$ and $D^*$
self-energies in a self-consistent manner. We find that the modifications of the mass and width of these resonances in
the nuclear medium strongly depend on the coupling to channels
with $D$, $D^*$ and nucleon content. Moreover, the resonances close to
the $DN$ or $D^*N$ thresholds change their properties more evidently
as compared to those far offshell.

The in-medium modifications of these resonant states have important consequences on the properties of open-charm mesons in matter. In fact, the properties of open-charm mesons in matter have been object of a recent theoretical interest due to the consequences for charmonium suppression. Furthermore, there have been speculations about the existence of $D$-meson bound states in nuclei resulting from a strong attractive potential felt by the $D$ mesons in nuclear matter \cite{Tsushima:1998ru}.  

The self-energy and, hence, spectral functions for $D$ and $D^*$ mesons are obtained self-consistently in a simultaneous manner, as it follows from HQSS. In order to solve the on-shell coupled-channel Bethe-Salpeter equation in matter, we take, as bare interaction, the extended WT interaction previously described and incorporate the in-medium modifications due to Pauli blocking in the intermediate nucleonic states and the open charm meson self-energies  \cite{Tolos:2009nn}. The self-energy for $D$ ($D^*$) meson is given by 
\begin{equation}
S(q^0,\vec{q}\,)
=-\frac{1}{\pi} \frac{{\rm Im} \Pi(q^0,\vec{q}\,)}{|(q^0)^2-\vec{q}\,^2-m^2-\Pi(q^0,\vec{q}\,)|^2} ,
\end{equation}
with $m$ the mass of the $D$ $(D^*)$ meson and $\Pi(q^0,\vec{q}\,)$ the $D$ $(D^*)$ self-energy, which depends on the energy $q^0$ and the momentum $\vec{q}$ of the $D$ $(D^*)$ meson.

The $D$ and $D^*$ spectral functions are displayed  for two densities at zero momentum on the l.h.s. of  Fig.~\ref{fig2}. Those spectral functions show a rich spectrum of resonance ($Y_c$) - hole ($N^{-1}$) states. On one hand, the $D$-meson quasiparticle peak mixes strongly with $\Sigma_c(2823)N^{-1}$ and $\Sigma_c(2868)N^{-1}$ states. On the other hand, the $\Lambda_c(2595)N^{-1}$ is clearly visible in the low-energy tail.  As for the $D^*$ meson, the $D^*$ spectral function incorporates the $J=3/2$ resonances, and the quasiparticle peak fully mixes with the  $\Sigma_c(2902)N^{-1}$ and $\Lambda_c(2941)N^{-1}$ states.  For both mesons, the $Y_cN^{-1}$ modes tend to smear out and the spectral functions broaden with increasing density. 

\section{Open-charm mesons in nuclei}

\begin{figure}[t]
\begin{center}
\includegraphics[width=0.3\textwidth,angle=-90]{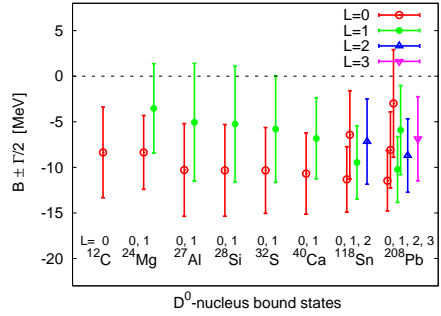}
\caption{$D^0$-nucleus bound states (taken from Ref.~\cite{GarciaRecio:2010vt}). \label{fig3}}
\end{center}
\end{figure}
 

\begin{figure}[t]
\begin{center}
\includegraphics[width=0.47\textwidth]{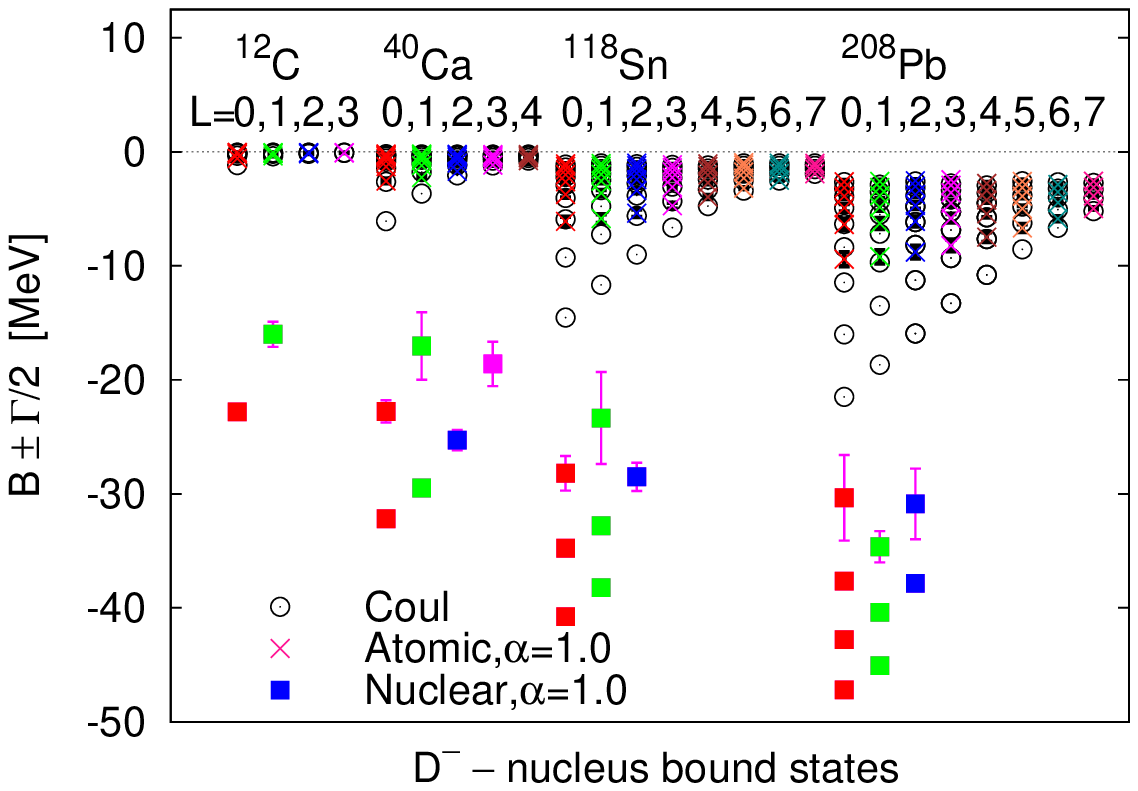}
\hfill
\includegraphics[width=0.47\textwidth]{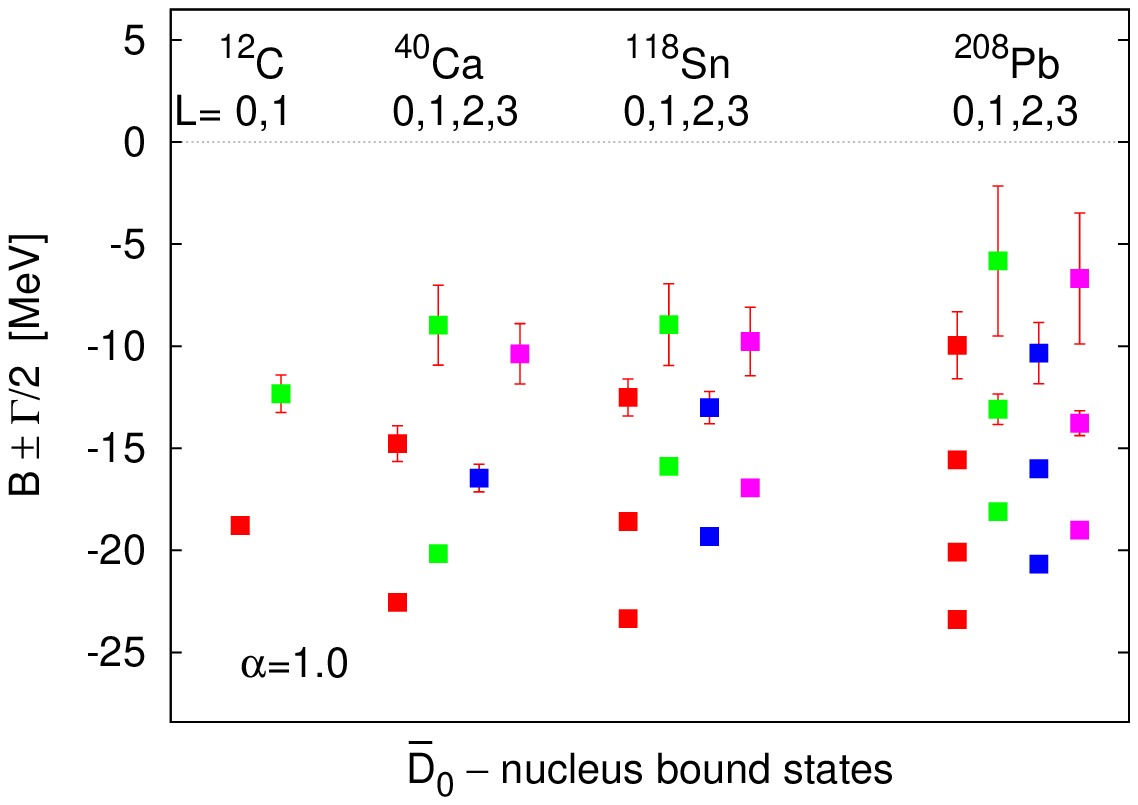}
\caption{$D^-$ and $\bar D^0$- nucleus bound states (taken from Ref.~\cite{GarciaRecio:2011xt}). \label{fig4}}
\end{center}
\end{figure}

$D$ and $\bar D$-meson bound states in $^{208}$Pb were predicted in Ref.~\cite{Tsushima:1998ru} using an attractive  $D$ and $\bar D$ -meson potential in the nuclear medium. This potential was obtained within a quark-meson coupling (QMC) model \cite{Guichon:1987jp}. The experimental observation of these bound states, though, might be problematic since, even if there are bound states, their widths could be very large compared to the separation of the levels. 

In order to explore the possible formation $D$-mesic nuclei within our model that respects HQSS, we solve the Schr\"odinger equation in the local density approximation. We use an energy dependent optical potential
\begin{equation}
  V(r,E) = \frac{
  \Pi(q^0=m+E,\vec{q}=0,~\rho(r))}{2 m},
\label{eq:UdepE}
\end{equation}
where $E=q^0-m$ is the $D$ or $\bar D$ energy excluding its mass, and $\Pi$ the meson self-energy. The optical potential resulting from our model is displayed for different densities on the r.h.s of Fig.~\ref{fig2}. For $D$ mesons we observe a strong energy dependence of the potential close to the $D$ meson mass due to the mixing of the quasiparticle peak with the  $\Sigma_c(2823)N^{-1}$ and $\Sigma_c(2868)N^{-1}$ states. As for the $\bar D$ meson, the presence of a bound state at 2805 MeV \cite{Gamermann:2010zz}, almost at $\bar D N$ threshold, makes the potential also strongly energy dependent. This is in contrast to the SU(4) model \cite{GarciaRecio:2010vt}.

Within our model, we obtain that the $D^0$-nucleus states are weakly bound (see Fig.~\ref{fig3}), in contrast to previous results using the QMC model. Moreover,  those states have significant widths \cite{GarciaRecio:2010vt}, in particular, for $^{208}$Pb. Only $D^0$-nucleus bound states are possible since the Coulomb interaction forbids the formation of bound states for $D^+$ mesons. It is also interesting to note that a recent work suggests the possibility that, for the lightest nucleus, $DNN$ develops a bound and narrow state with $S=0, I=1/2$ \cite{Bayar:2012dd}.

As for $\bar D$ mesons  in nuclei, not only $D^-$ but also $\bar{D}^0$ bind in nuclei  \cite{GarciaRecio:2011xt} (see Fig.~\ref{fig4}). The spectrum 
contains states of atomic and of nuclear types for all nuclei for $D^-$  while, as expected, only nuclear states are present for $\bar{D}^0$ in nuclei. Compared to the pure Coulomb levels, the atomic states are less bound. The nuclear ones are more bound and may present a sizable width. Moreover, nuclear states only exist for low angular momenta. 

The information on bound states is very valuable in order to gain some knowledge on the charmed meson-nucleus interaction. In fact,  this is of special interest  for the PANDA experiment at FAIR. However, the experimental observation of $D$ and $\bar D$-meson bound states is a difficult task since open-charm mesons with high momenta are produced in antiproton-nucleus collisions and it is a challenge to bind them in nuclei \cite{GarciaRecio:2010vt}.

\section{$D$-meson propagation in hot dense matter}

Another possibility to study the interaction of $D$ mesons with hadrons in matter is the analysis of transport coefficients for $D$ mesons in a hot dense medium composed of light mesons and baryons, such as it is formed in heavy-ion collisions. In fact, $D$ mesons are one of the cleanest probes of the early stages in heavy-ion collisions since their relaxation time is long enough so that they do not fully thermalize, but short enough to undergo significant reinteractions which reflect on their coupling with the medium.

The $D$-meson propagation can be studied using a kinetic description for the distribution function by means of solving the corresponding Fokker-Planck equation. The two relevant quantities to be determined are the drag ($F_i$) and diffusion coefficients ($\Gamma_{ij}$) of $D$ mesons in hot dense matter.  These are obtained from an effective field theory that incorporates both the chiral and HQSS in the meson \cite{Abreu:2011ic} and baryon sectors \cite{Tolos:2013kva}.

One interesting observable is the spatial diffusion coefficient $D_x$ that appears in Fick's diffusion law. Assuming an isotropic bath, this coefficient is given by 
\begin{equation}
\label{eq:dx} 
D_x= \lim_{p \rightarrow 0} \frac{\Gamma (p)}{m_D^2 F^2 (p)} \ , 
\end{equation}
in terms of the scalar $F(p)$ and $\Gamma(p)$ coefficients. The fact that previous works on viscous coefficients, such as the shear and bulk viscosities, show an extremum around the transition temperature between the hadronic and QGP phases makes the study of this coefficient around the transition temperature a topic of interest \cite{Tolos:2013kva}.

\begin{figure}[t]
\begin{center}
\includegraphics[width=0.4\textwidth,height=0.3\textwidth]{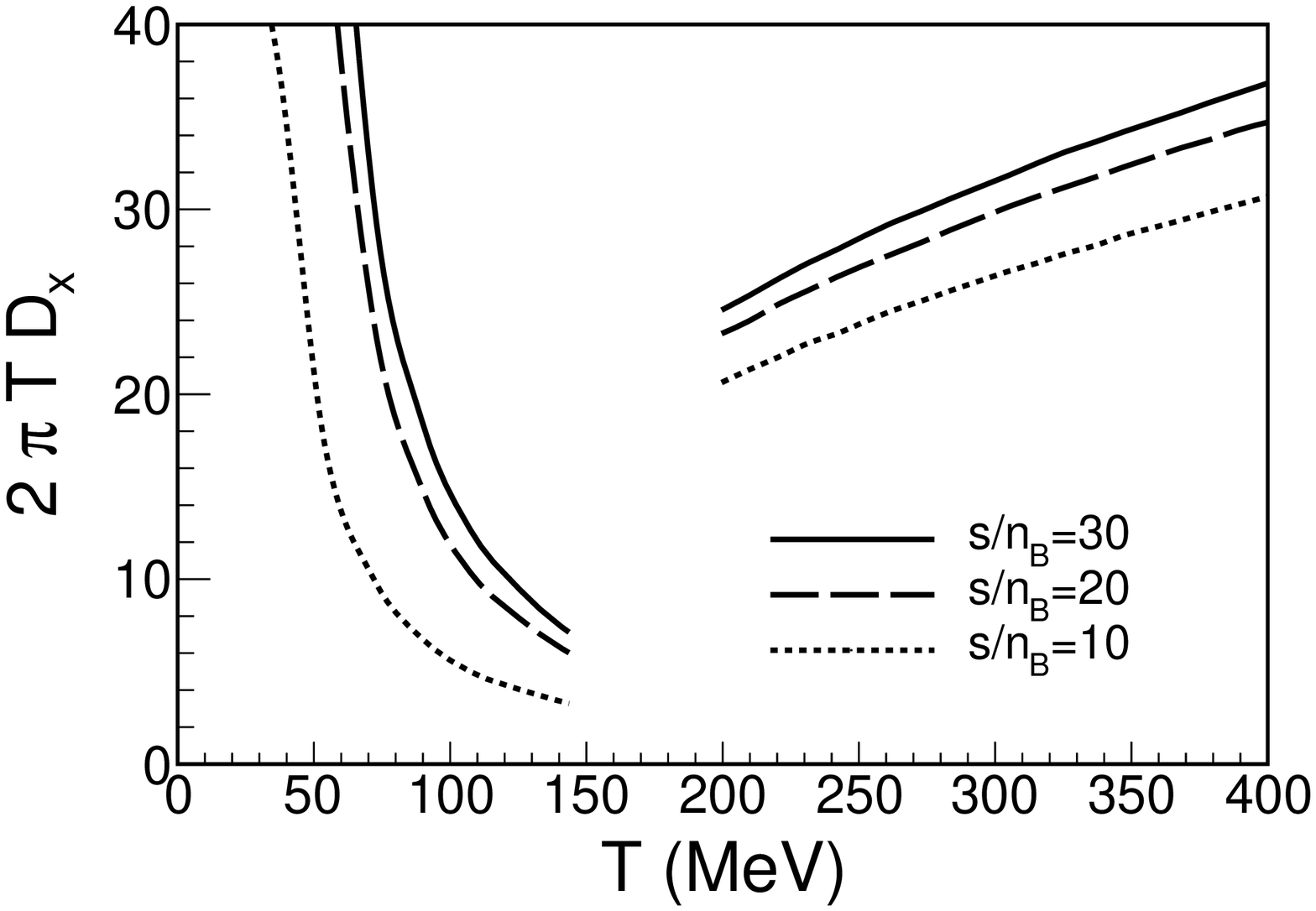}
\caption{
The coefficient $2\pi T D_x$ around the transition temperature (taken from Ref.~\cite{Tolos:2013kva}). For recent updates in the high-temperature phase, see Refs.~\cite{Berrehrah:2014tva,Ozvenchuk:2014rpa}. \label{fig5}}
\end{center}
\end{figure}

In Fig.~\ref{fig5}  we show  $2\pi T D_x$ around the transition temperature between the hadronic gas and the perturbative QCD calculation for QGP following isentropic trajectories ($s/n_B$=ct) from RHIC to FAIR energies. We observe that the dependence of the $2\pi T D_x$ on the entropy per baryon is similar in both phases: as long as
one increases the entropy per baryon (higher beam energies), the coefficient augments in both phases. The possible matching between curves in both phases for a
given $s/n_B$  seems to indicate the possible existence of a minimum in the $2\pi T D_x$ at the phase transition.  Indeed, a minimum at the phase transition following an isentropic trajectory from the hadronic to the QGP phase has been obtained very recently \cite{Berrehrah:2014tva,Ozvenchuk:2014rpa}.

\section*{Acknowledgments}
The author warmly thanks Carmen Garcia-Recio, Juan Nieves, Olena Romanets, Lorenzo Luis Salcedo and Juan Torres-Rincon for useful discussions and their continuous support that has made possible this work. This research was supported
by Ministerio de Economia y Competitividad under Contracts
No. FPA2010-16963 and No. FPA2013-43425-P,  from the Ramon y Cajal Research Programme of
Ministerio de Economia y Competitividad and from FP7-
PEOPLE-2011-CIG under Contract No. PCIG09-GA-2011-
291679.



\begin{thebibliography}{00}


\bibitem{lhc} http://home.web.cern.ch/topics/large-hadron-collider

\bibitem{fair} https://www.gsi.de/fair.htm

\bibitem{facility00} http://www.lepp.cornell.edu/Research/EPP/CLEO/; \\
http://belle.kek.jp/; \\
http://www.slac.stanford.edu/BF/

\bibitem{Tolos:2004yg}
  L.~Tolos, J.~Schaffner-Bielich and A.~Mishra,
  Phys.\ Rev.\  C {\bf 70}, 025203 (2004)

\bibitem{Tolos:2005ft}
  L.~Tolos, J.~Schaffner-Bielich and H.~Stoecker,
  Phys.\ Lett.\ B {\bf 635}, 85 (2006)


\bibitem{Lutz:2003jw}
  M.~F.~M.~Lutz and E.~E.~Kolomeitsev,
  Nucl.\ Phys.\  A {\bf 730}, 110 (2004)

\bibitem{Lutz:2005ip} 
  M.~F.~M.~Lutz and E.~E.~Kolomeitsev,
  Nucl.\ Phys.\ A {\bf 755}, 29 (2005)

\bibitem{Hofmann:2005sw} 
  J.~Hofmann and M.~F.~M.~Lutz,
  Nucl.\ Phys.\ A {\bf 763}, 90 (2005)

\bibitem{Hofmann:2006qx}
  J.~Hofmann and M.~F.~M.~Lutz,
  Nucl.\ Phys.\  A {\bf 776}, 17 (2006)

\bibitem{Lutz:2005vx} 
  M.~F.~M.~Lutz and C.~L.~Korpa,
  Phys.\ Lett.\ B {\bf 633}, 43 (2006)


\bibitem{Mizutani:2006vq} 
  T.~Mizutani and A.~Ramos,
  Phys.\ Rev.\ C {\bf 74}, 065201 (2006)


\bibitem{Tolos:2007vh}
  L.~Tolos, A.~Ramos and T.~Mizutani,
  Phys.\ Rev.\ C {\bf 77}, 015207 (2008) 



\bibitem{JimenezTejero:2009vq} 
  C.~E.~Jimenez-Tejero, A.~Ramos and I.~Vidana,
  Phys.\ Rev.\ C {\bf 80},  055206 (2009)

\bibitem{Haidenbauer:2007jq}
  J.~Haidenbauer, G.~Krein, U.~G.~Meissner and A.~Sibirtsev,
  Eur.\ Phys.\ J.\  A {\bf 33}, 107 (2007) 

\bibitem{Haidenbauer:2008ff} 
  J.~Haidenbauer, G.~Krein, U.~G.~Meissner and A.~Sibirtsev,
  Eur.\ Phys.\ J.\ A {\bf 37}, 55 (2008)

\bibitem{Haidenbauer:2010ch}
  J.~Haidenbauer, G.~Krein, U.~G.~Meissner and L.~Tolos,
  Eur. Phys. J A {\bf 47}, 18 (2011) 


\bibitem{Wu:2010jy}
  J.~-J.~Wu, R.~Molina, E.~Oset and B.~S.~Zou,
  Phys.\ Rev.\ Lett.\  {\bf105}, 232001(2010)

\bibitem{Wu:2010vk} 
  J.~J.~Wu, R.~Molina, E.~Oset and B.~S.~Zou,
  Phys.\ Rev.\ C {\bf 84}, 015202 (2011)

\bibitem{Wu:2012md} 
  J.~J.~Wu, T.-S.~H.~Lee and B.~S.~Zou,
  Phys.\ Rev.\ C {\bf 85}, 044002 (2012)

\bibitem{Oset:2012ap} 
  E.~Oset, A.~Ramos, E.~J.~Garzon, R.~Molina, L.~Tolos, C.~W.~Xiao, J.~J.~Wu and B.~S.~Zou,
  Int.\ J.\ Mod.\ Phys.\ E {\bf 21}, 1230011 (2012)

\bibitem{Liang:2014kra} 
  W.~H.~Liang, T.~Uchino, C.~W.~Xiao and E.~Oset,
  arXiv:1402.5293 [hep-ph]


\bibitem{Isgur:1989vq}
  N.~Isgur, M.~B.~Wise,
  Phys.\ Lett.\  B {\bf 232}, 113 (1989);
  M.~Neubert,
  Phys.\ Rept.\  {\bf 245}, 259 (1994);
  A.~V.~Manohar and M.~B.~Wise,
  Camb.\ Monogr.\ Part.\ Phys.\ Nucl.\ Phys.\ Cosmol.\  {\bf 10}, 1 (2000)


\bibitem{GarciaRecio:2008dp} 
  C.~Garcia-Recio, V.~K.~Magas, T.~Mizutani, J.~Nieves, A.~Ramos, L.~L.~Salcedo and L.~Tolos,
  Phys.\ Rev.\ D {\bf 79}, 054004 (2009) 

\bibitem{Gamermann:2010zz} 
  D.~Gamermann, C.~Garcia-Recio, J.~Nieves, L.~L.~Salcedo and L.~Tolos,
  Phys.\ Rev.\ D {\bf 81},  094016 (2010)

\bibitem{Romanets:2012hm} 
  O.~Romanets, L.~Tolos, C.~Garcia-Recio, J.~Nieves, L.~L.~Salcedo and R.~G.~E.~Timmermans,
  Phys.\ Rev.\ D {\bf 85}, 114032 (2012)

\bibitem{GarciaRecio:2012db} 
  C.~Garcia-Recio, J.~Nieves, O.~Romanets, L.~L.~Salcedo and L.~Tolos,
  Phys.\ Rev.\ D {\bf 87}, 034032 (2013)
    
\bibitem{Garcia-Recio:2013gaa} 
  C.~Garcia-Recio, J.~Nieves, O.~Romanets, L.~L.~Salcedo and L.~Tolos,
  Phys.\ Rev.\ D {\bf 87}, 074034 (2013)

\bibitem{Tolos:2013gta} 
  L.~Tolos,
  Int.\ J.\ Mod.\ Phys.\ E {\bf 22}, 1330027 (2013)

  
\bibitem{Gamermann:2011mq} 
  D.~Gamermann, C.~Garcia-Recio, J.~Nieves and L.~L.~Salcedo,
  Phys.\ Rev.\ D {\bf 84}, 056017 (2011) 

\bibitem{pdg}
K.A. Olive {\it et al.} (Particle Data Group), Chin. Phys. C {\bf 38}, 090001 (2014)

\bibitem{Jido:2003cb} 
  D.~Jido, J.~A.~Oller, E.~Oset, A.~Ramos and U.~G.~Meissner,
  Nucl.\ Phys.\ A {\bf 725}, 181 (2003)


\bibitem{Tsushima:1998ru} 
  K.~Tsushima, D.~H.~Lu, A.~W.~Thomas, K.~Saito and R.~H.~Landau,
  Phys.\ Rev.\ C {\bf 59}, 2824 (1999)
  


\bibitem{Tolos:2009nn} 
  L.~Tolos, C.~Garcia-Recio and J.~Nieves,
  Phys.\ Rev.\ C {\bf 80}, 065202 (2009)

\bibitem{Guichon:1987jp} 
  P.~A.~M.~Guichon,
  Phys.\ Lett.\ B {\bf 200}, 235 (1988)

\bibitem{GarciaRecio:2010vt} 
  C.~Garcia-Recio, J.~Nieves and L.~Tolos,
  Phys.\ Lett.\ B {\bf 690}, 369 (2010)


\bibitem{Bayar:2012dd} 
  M.~Bayar, C.~W.~Xiao, T.~Hyodo, A.~Dote, M.~Oka and E.~Oset,
  Phys.\ Rev.\ C {\bf 86}, 044004 (2012)


\bibitem{GarciaRecio:2011xt} 
  C.~Garcia-Recio, J.~Nieves, L.~L.~Salcedo and L.~Tolos,
  Phys.\ Rev.\ C {\bf 85}, 025203 (2012)
  
    
\bibitem{Abreu:2011ic} 
  L.~M.~Abreu, D.~Cabrera, F.~J.~Llanes-Estrada and J.~M.~Torres-Rincon,
  Annals Phys.\  {\bf 326}, 2737 (2011)
  

\bibitem{Tolos:2013kva} 
  L.~Tolos and J.~M.~Torres-Rincon,
  Phys.\ Rev.\ D {\bf 88}, 074019 (2013)
  
\bibitem{Berrehrah:2014tva} 
  H.~Berrehrah, P.~B.~Gossiaux, J.~Aichelin, W.~Cassing, J.~M.~Torres-Rincon and E.~Bratkovskaya,
  Phys.\ Rev.\ C {\bf 90},  051901 (2014)
  
\bibitem{Ozvenchuk:2014rpa} 
  V.~Ozvenchuk, J.~M.~Torres-Rincon, P.~B.~Gossiaux, L.~Tolos and J.~Aichelin,
  Phys.\ Rev.\ C {\bf 90},  054909 (2014)
%
%

\end{thebibliography}
%
%

\end{document}